%% file: ifip_conf.tex
\documentclass[conference,a4paper]{IEEEtran}
\IEEEoverridecommandlockouts
 
\usepackage{geometry}
\usepackage[draft]{hyperref}
\usepackage{orcidlink}
\usepackage{cite,flushend}
\usepackage{graphicx}
\usepackage{amsmath,amssymb,amsfonts}
\usepackage{amssymb} 

\usepackage{graphicx}
\usepackage{textcomp}
\usepackage{xcolor}
\usepackage{bbm,bm}
\usepackage{svg}
\usepackage[ruled]{algorithm}
\usepackage[algo2e]{algorithm2e} 
\usepackage{algorithmic}
\usepackage{ragged2e}
\usepackage[nonumberlist, style=long, toc, acronym, symbols]{glossaries}
\usepackage{orcidlink}
\usepackage{multirow}
\usepackage{colortbl}
\usepackage{tabulary}
\usepackage{etoolbox}
\usepackage{tcolorbox}
\usepackage{empheq}
\usepackage[normalem]{ulem}
\usepackage{tabularx} 
\usepackage{booktabs} 
\usepackage{comment}
\usepackage{setspace}
\usepackage{color}
\usepackage{tabularray}
\definecolor{JapaneseLaurel}{rgb}{0,0.501,0}

\begin{document}
\include{glossary}
\title{Towards End-to-End Network Intent Management with Large Language Models}
%
%
%
%

\author{Lam~Dinh, Sihem Cherrared, Xiaofeng Huang, and Fabrice Guillemin  \\
        \IEEEauthorblockA{ \textit{Orange Innovation},  France  \\
}

}

\IEEEtitleabstractindextext{%
\begin{abstract}

\acrfull{llms} are likely to play a key role in \acrfull{ibn} as they show remarkable performance in interpreting human language as well as code generation, enabling the translation of  high-level intents expressed by humans into low-level network configurations. In this paper, we leverage closed-source language models (i.e., Google Gemini 1.5 pro, ChatGPT-4) and open-source models (i.e., LLama, Mistral) to investigate their capacity to generate \acrshort{e2e} network configurations for radio access networks (RANs) and core networks in 5G/6G mobile networks. We introduce a novel performance metrics, known as FEACI, to quantitatively assess the  format (F), explainability (E), accuracy (A), cost (C), and inference time (I)  of the generated answer; existing general metrics are unable to capture these features. The results of our study demonstrate that open-source models can achieve comparable or even superior translation performance compared with the closed-source models requiring costly hardware setup and not accessible to all users.
\end{abstract}

}

\maketitle

\IEEEdisplaynontitleabstractindextext

%
\IEEEpeerreviewmaketitle

\section{Introduction}\label{sec:introduction}

%
%
%
%
The complexity of network services and applications has  continually grown with the advent of novel technologies, especially in the recent few years with the emergence of virtualization technologies applied to  5G networks (e.g., containerized network functions, specialized or shared  network slices, etc.). While integrating new network services is necessary to enhance network performance and facilitates greater variety of applications, it inevitably creates difficulties for network management because network configurations, which  contain a large number of technical parameters in different network domains, are no longer sustainable for human operation \cite{abbasNetworkSliceLifecycle2021}. This requires new network management strategies, notably Zero-Touch Service Management (ZSM), in which the network operations are fully automated through \textit{self-healing}, \textit{self-configuration}, and \textit{self-optimization} \cite{GSZSM016}.

Policy-driven and intent-based mechanisms are among the key concepts that enable autonomous networks, as highlighted in \cite{liyanageSurveyZeroTouch2022}. Specifically, ZSM utilizes the notions of \acrfull{ibn} \cite{clemmIntentBasedNetworkingConcepts2022} for enabling end-users to specify their high-level network operational goals by defining intents through various  network interfaces. This method fills the technical gap between novice users and network management, as \acrshort{ibn} allows users with minimal network expertise to focus on the desired network operational goals, rather than on technical specifications. 

Once the user  has well defined intents, \acrshort{ibn} plays an important role in translating them into technical policy configurations and actions, that can be consumed by the available set of controllers and orchestrators for further deployment. For instance, any non-expert user can interact with network  to establish both 5G \acrshort{embb} slice and \acrshort{urllc} slice for a particular region, using natural language descriptions as follows: \textit{``I need to setup \acrshort{embb} and \acrshort{urllc} slices for the Paris region. For the \acrshort{embb} slice, our service covers up to 10,000 users at the same time, with a minimum throughput of 100 Mbps. For the \acrshort{urllc} slice, we require to meet a maximum end-to-end latency of 10ms in at least 99,999 $\%$ of time for a minimum of 1000 users."}. To translate this description into a deployable network  configuration, \acrshort{ibn} first tries to extract network intents (i.e., \acrshort{embb}, 10000 users, 100 Mbps; \acrshort{urllc} 1000 users, 10ms of delay, etc.) before creating detailed technical configurations of network functions (i.e., RAN and Core Network functions) for the building  and running phases. The entire process is carried-out via intent life-cycle management procedures, which involves five stages: \textit{profiling}, \textit{translation}, \textit{resolution}, \textit{activation}, and \textit{assurance} \cite{leivadeasSurveyIntentBasedNetworking2023}.

However, there are multiple challenges associated with the use of \acrshort{ibn} for \acrfull{e2e} network management: \textit{(i)} High-level network intents must be understandable to both the user and the network. For this reason, they are typically documented in human-readable languages (i.e., JSON, YAML, etc.), and are formularized by certain standards (i.e., 3GPP, TMForum, etc.) with a large number of technical requirements, subject to errors. Therefore, users must master such description methods to harness the full potential of the intent concept. Sometimes, it is challenging for non-expert users to manually describe all the network objectives in simple natural language, which implies additional complexities for network management.  In addition,  when the technical intents are fully extracted, it is difficult to  map intents onto specific network functions during the network activation phase. 

All those difficulties related to intent life-cycle management can potentially be mitigated with \acrfull{llms} thanks to their capabilities of understanding, generating, and interpreting human language and facilitating network management for users without extensive technical expertise ~\cite{liuLargeLanguageModels2024,maatoukLargeLanguageModels2024}. However, the practical applications of \acrshort{llm} for end-to-end network management is still limited, because it lacks to some extent  performance indicators that measure how good/bad the obtained answers can be used for service deployments.  Furthermore, most of the studies focused on closed-source GPT models (e.g., GPT-4), which restrict our understanding of what occurs within this black-box.  

In the objective of developing an  \acrshort{llm} based \acrshort{ibn} framework with verifiable output, our key contributions are  as follows:
\begin{itemize}
    \item First, we define an \acrshort{ibn} framework, in which business intent resolver and service intent resolver are introduced in the intent layer for translating high level user expressions into network resource configurations. This architecture is compatible with the  Open Digital Architecture (ODA) specified by the TM Forum.   
    \item Second,  we improve the responses of \acrshort{llm} with regard to  intent translation. In this study, we apply prompting on both closed-source and open-source \acrshort{llm} models to study the generated network configurations.  
    \item Finally, we propose a new evaluation benchmark, referred to  as  FEACI, to objectively check the generated responses from \acrshort{llm} models against the expectation of the network providers in 5 factors: Format (F), Explainability (E), Accuracy (A), Cost (C), and Inference time (I). 
\end{itemize}

The paper is organized as follows: Related work is discussed in Section~\ref{sec:related_work}. Section~\ref{sec:solution} presents the general system model and our contribution, whereas Section~\ref{sec:results} provides numerical results. Finally, Section~\ref{sec:conclusion} presents concluding remarks.

\section{Related works}\label{sec:related_work}

In the literature,  \acrfull{nlp}  interfaces between users and the network  to facilitate network management through the use of natural language has been considered in various studies, see for instance \cite{mcnamaraNLPPoweredIntent2023,jacobsHeyLumiUsing2021}.  This approach has however several shortcomings when accomplishing  complex tasks that require higher language understanding and generation capabilities. Alternatively, \acrfull{llms} have been proposed  to interpret and transform user intents into network policies and/or generate network configurations to match user descriptions \cite{brodimasIntentbasedNetworkManagement2024, orlandiIntentBasedNetworkManagement2024, maniasIntentBasedNetworkManagement2024}. 
The use of \acrshort{nlp}/\acrshort{llms} in intent-based networking can be categorized into several topics as detailed in the following sections.

\subsection{Intent extraction} Mcnamara et al. \cite{mcnamaraNLPPoweredIntent2023} embed \acrshort{nlp} interface to the intent engine for facilitating the interaction between humans and network management systems for slice provisioning, indoor positioning, and network service deployment. The use of \acrshort{llms} is not considered in this work, and the direct mapping between user intents to the API endpoint is still missing. 

 Orlandi et al. \cite{orlandiIntentBasedNetworkManagement2024} illustrates a framework that allows translation of network intents into 5G services, without the need for telco expertise. Empowered by \acrshort{nlp}, this Proof-of-Concept (PoC) introduces two level of resolvers (i.e., \textit{Business Intent Resolver} and \textit{Service Intent Resolver}), which automatically capture the network intents from the non-expert text input and continuously verify the deployment status of ordered services.  
 
 Cesila et al. \cite{cesilaChatIBNRASABuildingIntent2023} introduces a promising integration of an open-source RASA Chatbot into the Intent (Translator) Engine, helping users configure the optical network through high-level language communication. Although this work sets up  a framework that allows a high level of customization, human-in-the-loop is required to validate the responses of the chatbot and limits the scalability of the solution.
 
Manias et al. \cite{maniasIntentBasedNetworkManagement2024} demonstrates the potential use of \acrshort{llms} for intent extraction of core network operations in 5G and beyond network.  Based on prompts that include task description and background contexts, closed-source \acrshort{llms} in this work provide the extraction of network intents in the response together with explainability. However, the evaluation is carried out on  closed-source \acrshort{llm} models (i.e., GPT-4), while open-source models are not considered for further examination.  
Besides natural language intents, \acrshort{llms} have also been used to extract meaningful user intents from log data \cite{shahUsingLargeLanguage2024} \cite{carrionTaxonomyGenerationTool2019}.

\subsection{Intent activation/assurance} Once the intents are translated from user demands into network requirements and policies, it is also crucial for the network orchestrator to deploy correct decision model to meet the service requirement. In this context,  Collet et al. \cite{colletLossLeaPLearningPredict2022} propose a forecasting model, so called \textit{LossLeap},  to anticipate network configurations and cope with complex objectives represented by the intents. 

The papers \cite{linAppleSeedIntentBasedMultiDomain2023, wangNetworkMeetsChatGPT2023} propose intent-based infrastructure management systems that leverage few-shot prompting in generic \acrshort{llm}s to translate natural language intents into computer programs for multi-domain infrastructure. However, their use of closed-source GPT-based model can be a bottleneck. Mekrache et al. \cite{mekracheIntentBasedManagementNextGeneration2024} address full Life-Cycle (LC) of intent management framework \cite{clemmIntentBasedNetworkingConcepts2022}, in which \acrshort{llms} play a centric role to configure and manage \acrshort{e2e} network services. Via the combination of  prompting and open-source \acrshort{llms} (e.g., Mistral, LLama, etc), they are capable of formatting Cloud/Edge intents and \acrshort{ran} intents from user description, and those decomposed configurations into JSON (e.g., slicing, latency, throughput, etc.) can be passed to the \acrfull{nfvo} for network service deployment. However, there are several drawbacks encountered in this work: \textit{(i)} the efficiency of intent extraction and translation are not fully examined, when it is not clear  if the \acrshort{llms} are able to produce efficient translations and it may lead to the conflicts in the intent activation, \textit{(ii)} Human Feedback is involved in the process, and it can be time-consuming.

\subsection{Our approach}
In this work, we apply both closed-source and open-source \acrshort{llms} to deal with (1) the translation  of high-level intents from users into low-level network configurations, and (2) the mapping between technical intents and network configurations, via prompting \cite{weiChainofThoughtPromptingElicits2023}. Furthermore, we introduce a novel performance metrics, namely FEACI (see the Introduction).

\section{End-to-End Network Intent Management}\label{sec:solution}
\subsection{Network Intent Architecture}

 \acrshort{ibn}  introduces  an additional  layer in addition to the business  and the network layers to manage the life-cycle of high level intents from their initiation at the business layer to practical realization at the network layer. The intent layer generally  achieves  five main functional blocks: \cite{leivadeasSurveyIntentBasedNetworking2023}: \textit{(i)} \textbf{Intent Profiling} is the communication gateway between IBN system and users. It is designed so that users can express their meaningful intents in a human-friendly way (i.e., through natural language expression, drop down menu, etc.). \textit{(ii)} \textbf{Intent Translation} is responsible for translating those abstracted intents into network policy that will be seen as low-level network configuration for network infrastructure. \textit{(iii)} \textbf{Intent Resolution} is used to resolve the conflicting intents that are infeasible in  current network states. \textit{(iv)} \textbf{Intent Activation} deploys feasible intents so that the networks are configured as intended by the user, and \textit{(iv)} \textbf{Intent Assurance} ensures that the network complies with the user intents throughout its lifetime.     

\begin{figure}[htp!]
    \centering
    \includegraphics[scale=0.5]{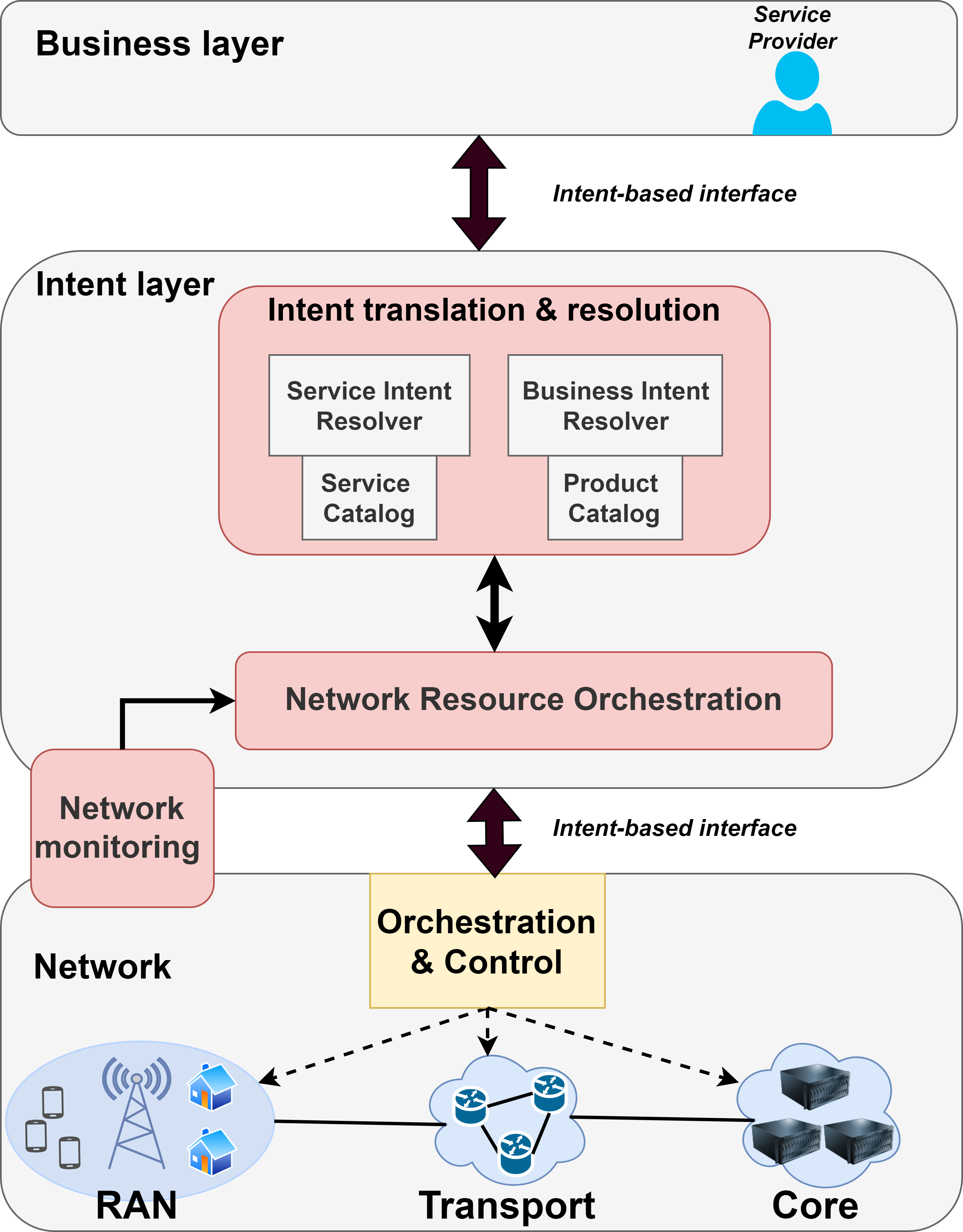}
    \caption{Network architecture  with Intent layer }
    \label{fig:ibn_archi}
\end{figure}

In this paper, we primarily focus on the the life-cycle management of an intent from the moment it is initiated till the low-level configuration of network function (i.e., Intent Profiling, Translation and Resolution.). The IBN architecture, which highlights our contributions by ignoring the not relevant functional blocks, is shown in Figure~\ref{fig:ibn_archi}. 

In the intent layer, the block responsible for intent translation  is  near the user interface to facilitate interactions with users and to gather user intents through natural language communications. There are two primary ways for users to convey their high-level objectives for network management: 
\begin{enumerate}
    \item Users \textbf{explicitly} describe network goals via natural language, for instance \textit{``I need to set up URLLC
slices for the Paris region. We aim to meet
a maximum end-to-end latency of 10ms in at least 99,999
$\%$ of time for a minimum of 1000 users."}. The intent translation and resolution block then extracts technical intents from user inputs (e.g., end-to-end latency of 10 ms, reliability: 99,999
$\%$, supported users: 1000 etc.), prior to parameterizing network functions to meet these goals. 
\item Users  specify \textbf{which/where/when} service they want to set up. This demand will be mapped into a product catalog that contains numerous services, each of which has been configured with technical parameters.  For example, user demands following this method can be \textit{``I need to establish URLLC slices for the Paris region"}. Afterwards, both ``URLLC service" (\textbf{which}) and ``Paris region" (\textbf{where}) from user demand are used to match the URLLC product, which is one of the products in the catalog, and contains network parameters such as end-to-end latency of 10 ms, reliability: 99,999 $\%$, supported users (1000), and so on.  
\end{enumerate}

\begin{figure}[htp!]
    \centering
    \includegraphics[scale=0.55]{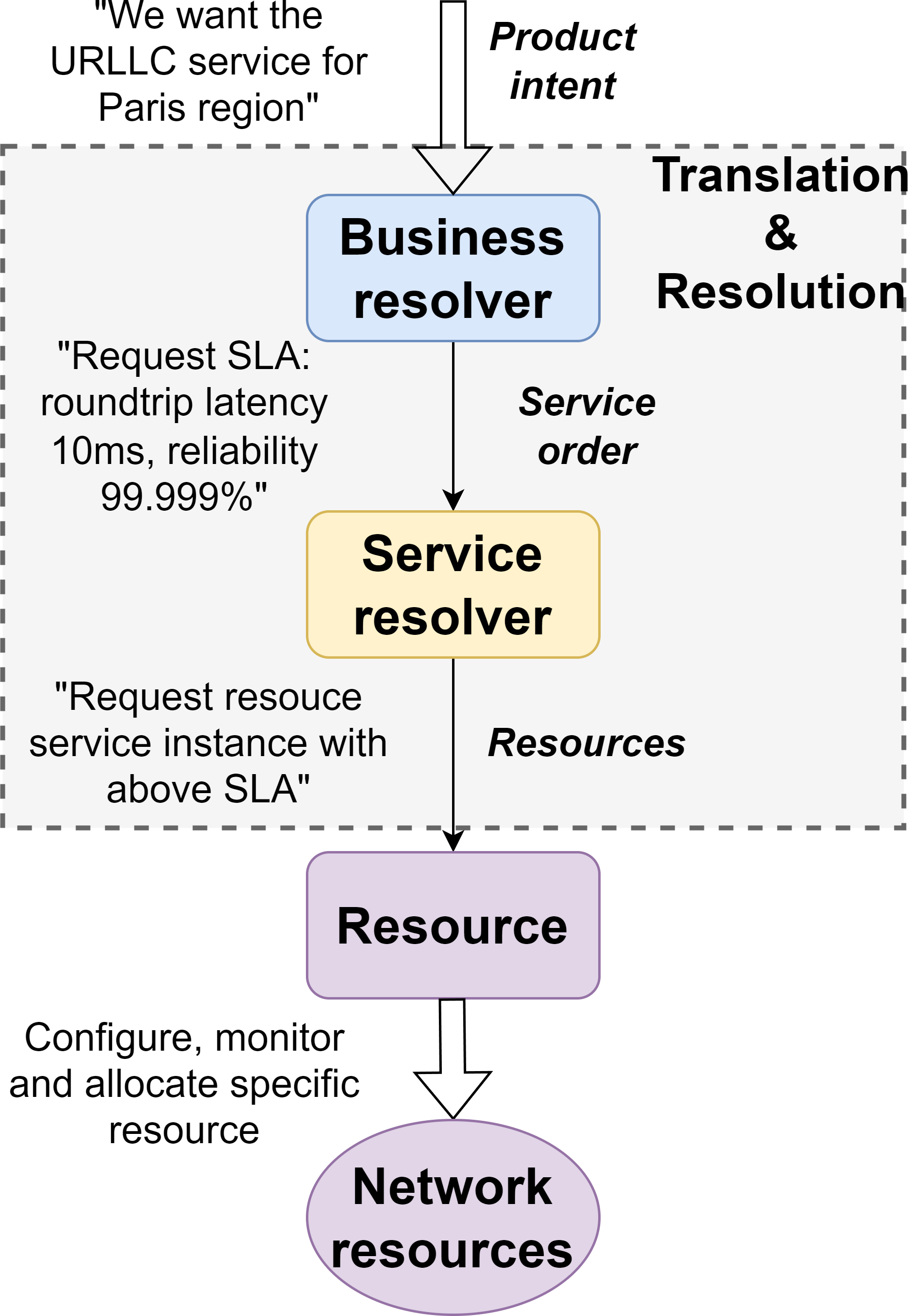}
    \caption{Translation and resolution block with business and service resolver }
    \label{fig:ibn_intent}
\end{figure}

In this paper, we target the second method within the intent translation and resolution block, as it enables non-expert users to express their needs without being network experts.  \textbf{Business Intent Resolver} and \textbf{Service Intent Resolver} are then used \cite{orlandiIntentBasedNetworkManagement2024} to handle the service orders. As illustrated in Figure~\ref{fig:ibn_intent}, the business resolver  translates the requirements from business users (i.e., product intent) into \textit{service order} specifying which specific \acrshort{sla}s are used to describe the user demand. Subsequently, service resolver transforms the service order into \textit{resource order} that can be orchestrated by the network resource orchestrator.

The order from business users is mapped onto a single intent-driven product, which contains a subset of (networking) parameters, such as expected E2E latency, supported users, deployment region, etc. For instance, any mission critical product can be characterized with low latency properties (i.e., in the range [1ms;10ms] for industrial use cases), or a high broadband product focus on higher network throughput services (i.e., [100Mpbs:1000Mbps] for consumer use cases). Each product may also be enriched with contextual and meta data information (e.g., ranking of user expertise for data exposure during service implementation) to better adapt to the expert or non-expert user profile. 

Once the product order is validated at the business resolver level, the \textbf{Service Intent Resolver} begins to verify if the technical intents described in the ordered product can be deployed. Subsequently, those technical intents are converted into technical solutions in vertical network domain (RAN, Transport, Edge/cloud, etc.) to achieve operational goals described in the product specifications. For example, those technical solutions are related to the  required cloud resources (e.g., CPU speed, RAM volume, storage, etc) for the core network, topology settings for transport network or radio-related configurations for RAN network. Note that those technical details are not exposed to the users, and are specified in the service catalog. 

Finally, the technical solutions, which are produced by the Service Intent Resolver, are handled by the  Network Resource Orchestrator (NRO) for their implementation. To ensure user satisfaction over the life-cycle of the intents, the NRO should implement proactive resource provisioning to accommodate the dynamic changes of the underlying network. For, it should continually  measures network status  and dynamically fine-tune the network configurations to meet user expectation. This point is not addressed in this paper.
In the following, we propose an approach that is based on \acrshort{llms} to manage  business and service resolver for interpreting user intents into RAN and Core configurations. 

\subsection{Large Language Models}

\acrshort{llms} such as GPT4 \cite{openaiGPT4TechnicalReport2024}, Gemini \cite{teamGeminiFamilyHighly2024}, etc.  are capable of understanding contextual data (i.e., texts, images, etc.) across a wide range of technical domain (i.e., telecom-specific, code generation, etc.). The primary factor behind the success of \acrshort{llm}s  is largely due to the transformer architecture \cite{vaswaniAttentionAllYou2023} that enables them to comprehend information context more effectively by  selectively paying attention to various parts of the input, and their ability to access a vast corpus of data (e.g., books, internet sources, etc. ) during training. Specifically, based on self-attention mechanism, which stands as a core component of transformers, the long-range dependencies and contextual relationships can be inferred from the input data. As a result, the generated output will be generated with high level of coherence and contextual relevance. This approach enhances the results compared to the use of \acrshort{rnn} as all relevant information is ephemeral and only represented by the current hidden state.   

\begin{figure}[htp!]
    \centering
    \includegraphics[scale=0.6]{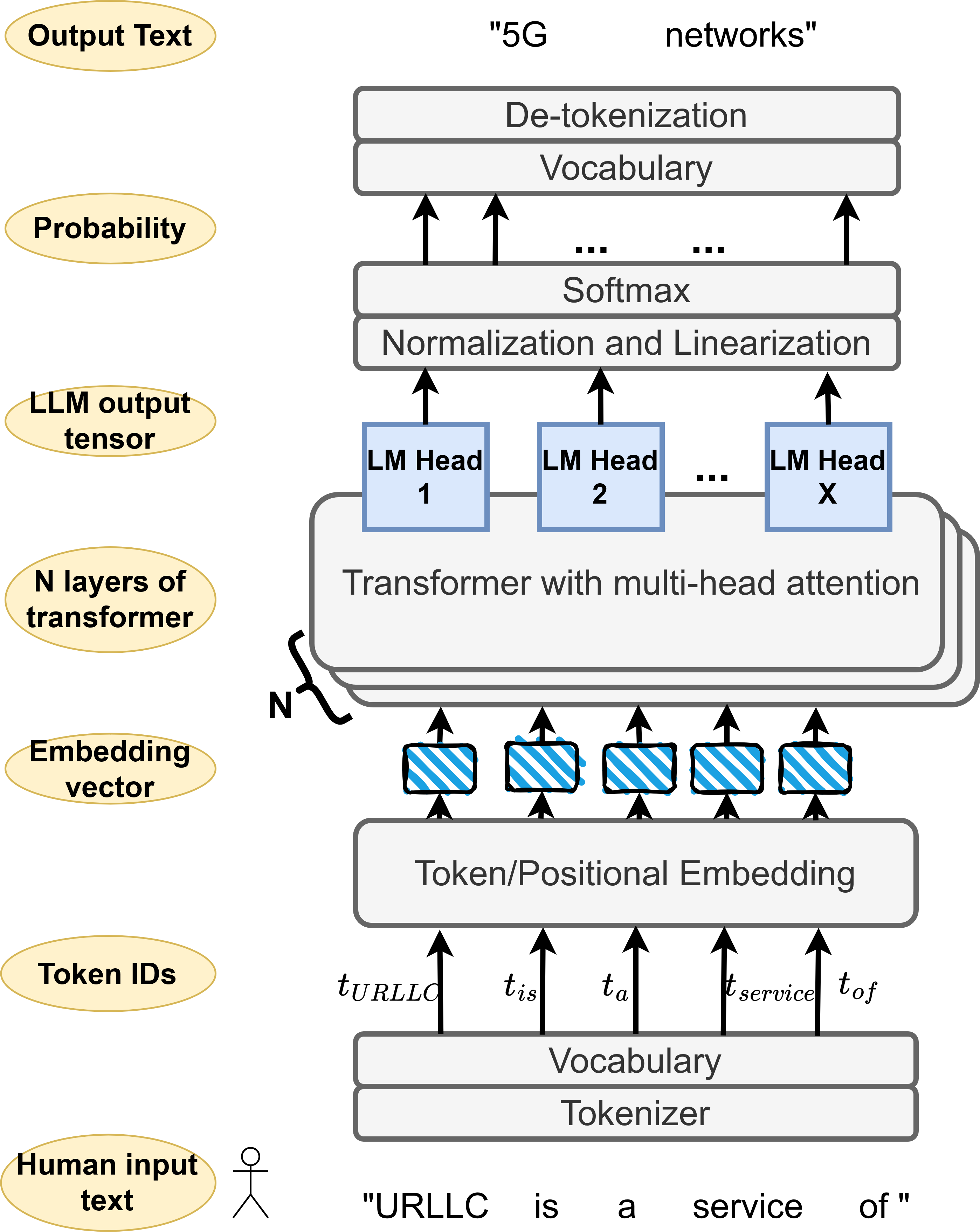}
    \caption{Illustration of an \acrshort{llm} workflow }
    \label{fig:llm}
\end{figure}

To demonstrate the relevance of an \acrshort{llm} in the \acrshort{ibn} context, Figure \ref{fig:llm} presents an \acrshort{llm} architecture, which is composed of several processing blocks: \textit{(i) Pre-processing}, \textit{(ii)} Transformer and \textit{(iii) Post-processing} . At the \textit{pre-processing} stage, human-understandable data is de-serialized into machine-understandable data using \textit{tokenization} and \textit{embedding}. They are responsible for converting human input sequence into corresponding trainable embedding vectors.

Inside the \textit{transformer} block, multi-head attention (i.e., LM head) with trainable weights is introduced to tackle the input sequences that have complex pattern recognition. In this case, each LM head is dedicated to learn different kinds of semantic information of input sequence. Furthermore, it is common in the \acrshort{llm} models that  a transformer is placed on top of another to iteratively apply the multi-head self-attention mechanism $N$ times in order to further enhance the efficiency of the input pattern recognition. Finally, the output of each LM head in the final layer of the transformer block is normalized before passing through a Feed-forward Neural Network (FNN) at the \textit{post-processing} step. The goal of this step is to map the \acrshort{llm} output tensors, which are processed within the transformer,  with the corresponding tokens to generate the response.


\subsection{From natural language to network deployment with \acrshort{llm}s}

\begin{figure*}[htp!]
    \centering
    \includegraphics[scale=0.6]{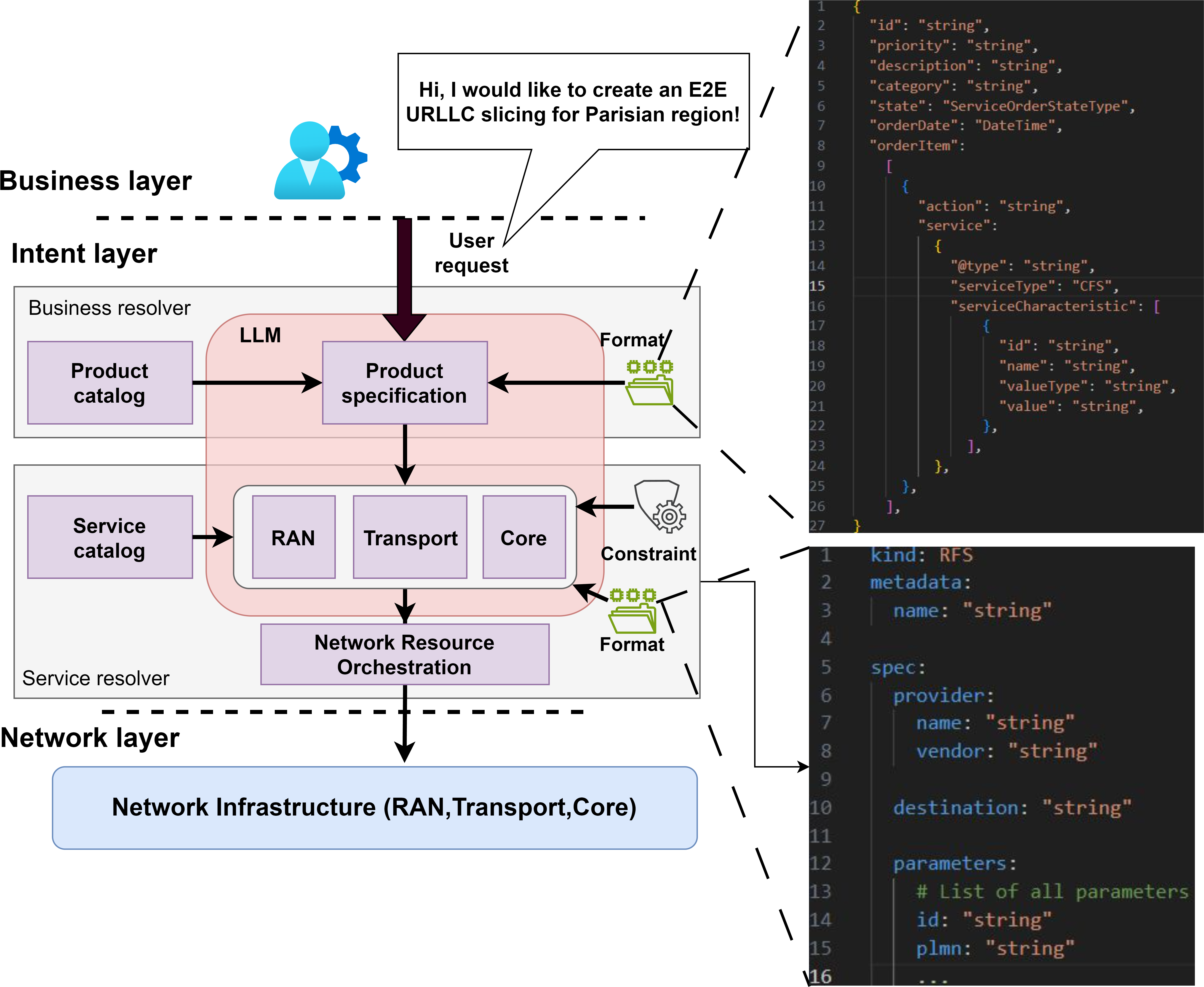}
    \caption{Intent translation and resolution with \acrlong{llms} }
    \label{fig:business_to_intents}
\end{figure*}

Figure \ref{fig:business_to_intents} shows our proposal with the use of \acrshort{llms} to translate user requests into operational intents, and to activate the corresponding network resources to meet user intents. In this architecture, users first interact with a chatbot based on \acrshort{llm} using natural language to express their needs, for instance \textit{``I would like to create an E2E \acrshort{urllc} slicing for Parisian region."}. Then, given the capability of \acrshort{llms} to understand human languages and semantic information, they manage to match user demands with \textit{at least} one of the network products that are available. 

It should be highlighted that each product contains multiple parameters of the desired services. In the business resolver, the product descriptions, which are represented as the operational intents, are usually formatted according to the TM Forum ODA definitions. It takes into account the general parameters of the creation order (e.g., required latency value, etc. ) as well as their metadata (e.g., $id$ in the product catalog, date of creation, category, etc.). This format is also known as \acrfull{cfs} specification. 

The second translation is also carried out by the \acrshort{llms} at the level of service resolver. Given the constraints of the service prior to the deployment, the available services in the catalog, and the TM Forum ODA format, a set of technical solutions, also known as \acrfull{rfs}, is produced from the \acrshort{cfs} description that is available in the first step. Those \acrshort{rfs} are formatted and ready to be fed onto the \acrfull{nro} for the imminent deployment.

\subsection{Prompting}

To better customize the output of the \acrshort{llm} models to match the user expectation, Chain-of-Thought (CoT) prompting is considered as an efficient approach~\cite{weiChainofThoughtPromptingElicits2023}. By providing the models with additional information via prompting, it improves the ability of the language models to understand similar tasks and efficiently respond to them. Practically, a CoT prompt includes a tuple $<Q_i, CoT_i,A_i>$, where $Q_i$ is a description of a task,  $CoT_i$ is a series of reasoning steps that lead to the result $A_i$. In this study, we consider zero-shot prompting (ZERO), one-shot prompting (ONE), and few-shot CoT prompting (FEW), which respectively provide zero example, one example of expected format output, and few examples of how to generate final output.  

Without providing any additional context to the \acrshort{llm} models, we examine in ZERO how they can generate technical intents in the "standard" format (i.e., 3GPP and TMF), of which they might not know during training. The generated answer from \acrshort{llm} models will be investigated in: (1) how good the format is, (2) how relevant of each technical intent is and (3) how feasible the models can do reasoning. In this regard, we create a target configuration, which serves as a reference for the performance comparison.

\begin{figure}[htp!]
    \centering
    \includegraphics[scale=0.55]{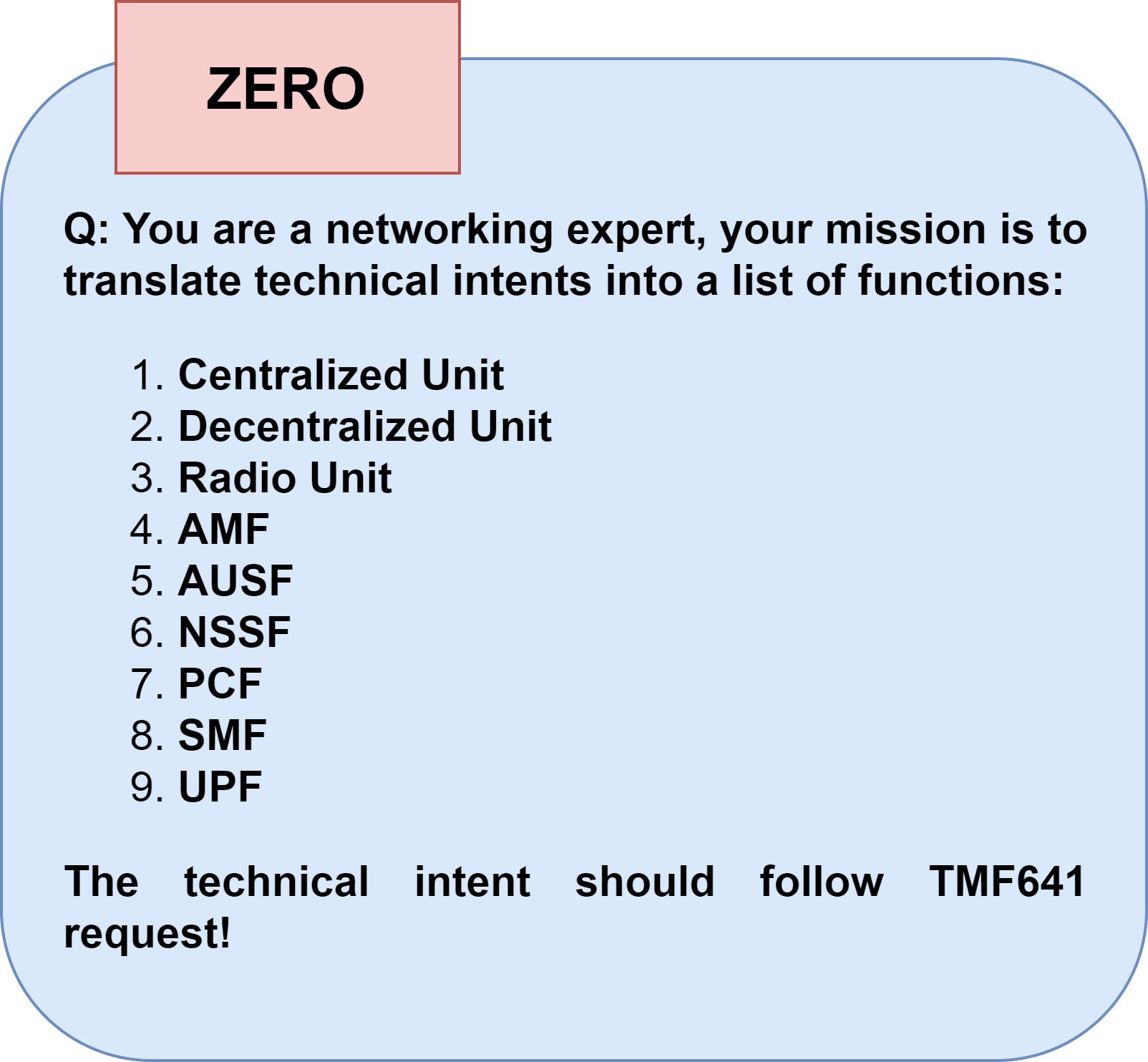}
    \caption{Zero-shot prompting (ZERO) }
    \label{fig:prompt0}
\end{figure}

\begin{figure}[t]
    \centering
    \includegraphics[scale=0.55]{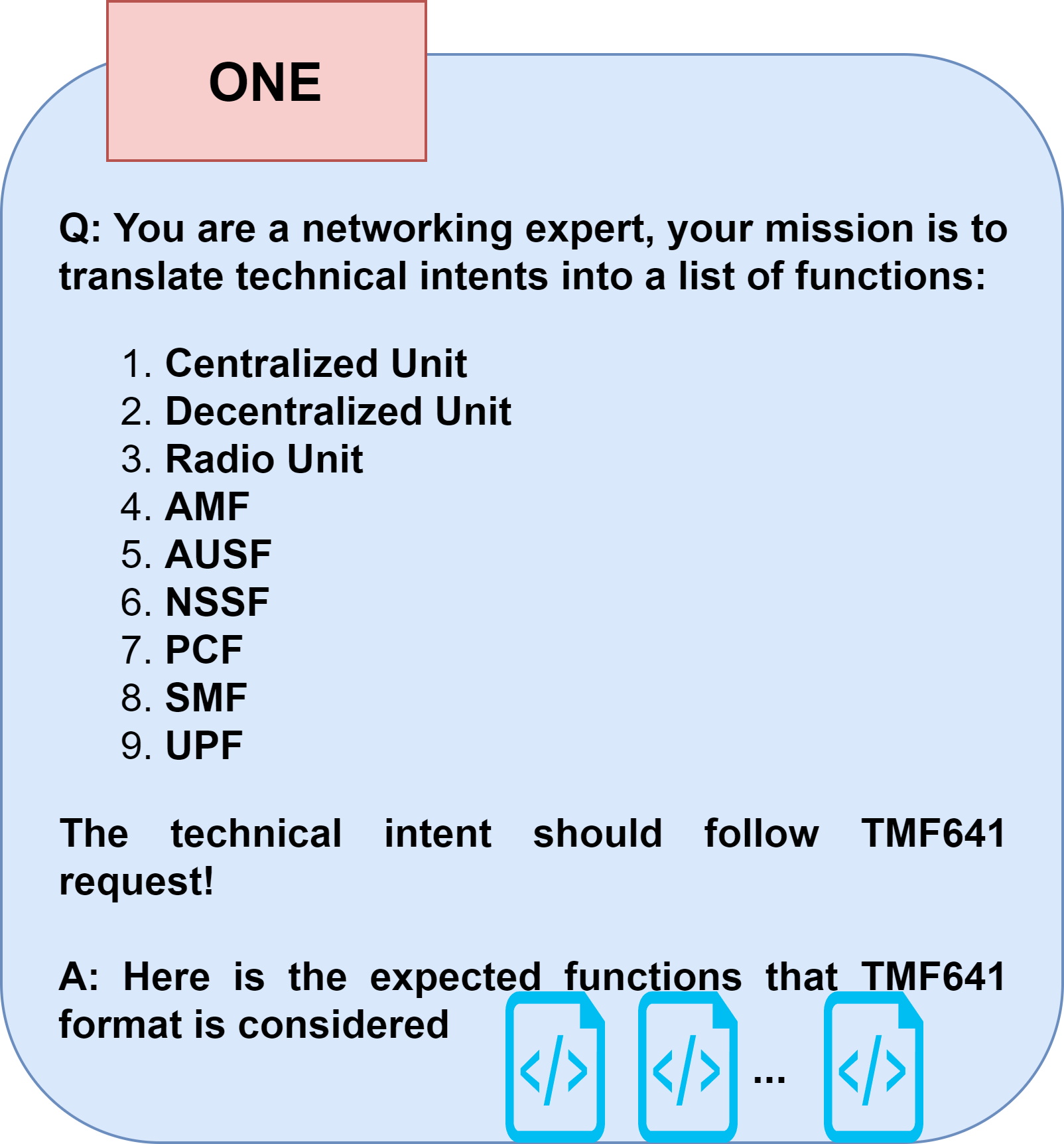}
    \caption{One-shot prompting (ONE) }
    \label{fig:prompt1}
\end{figure}

\begin{figure}[hbtp!]
    \centering
    \includegraphics[scale=0.55]{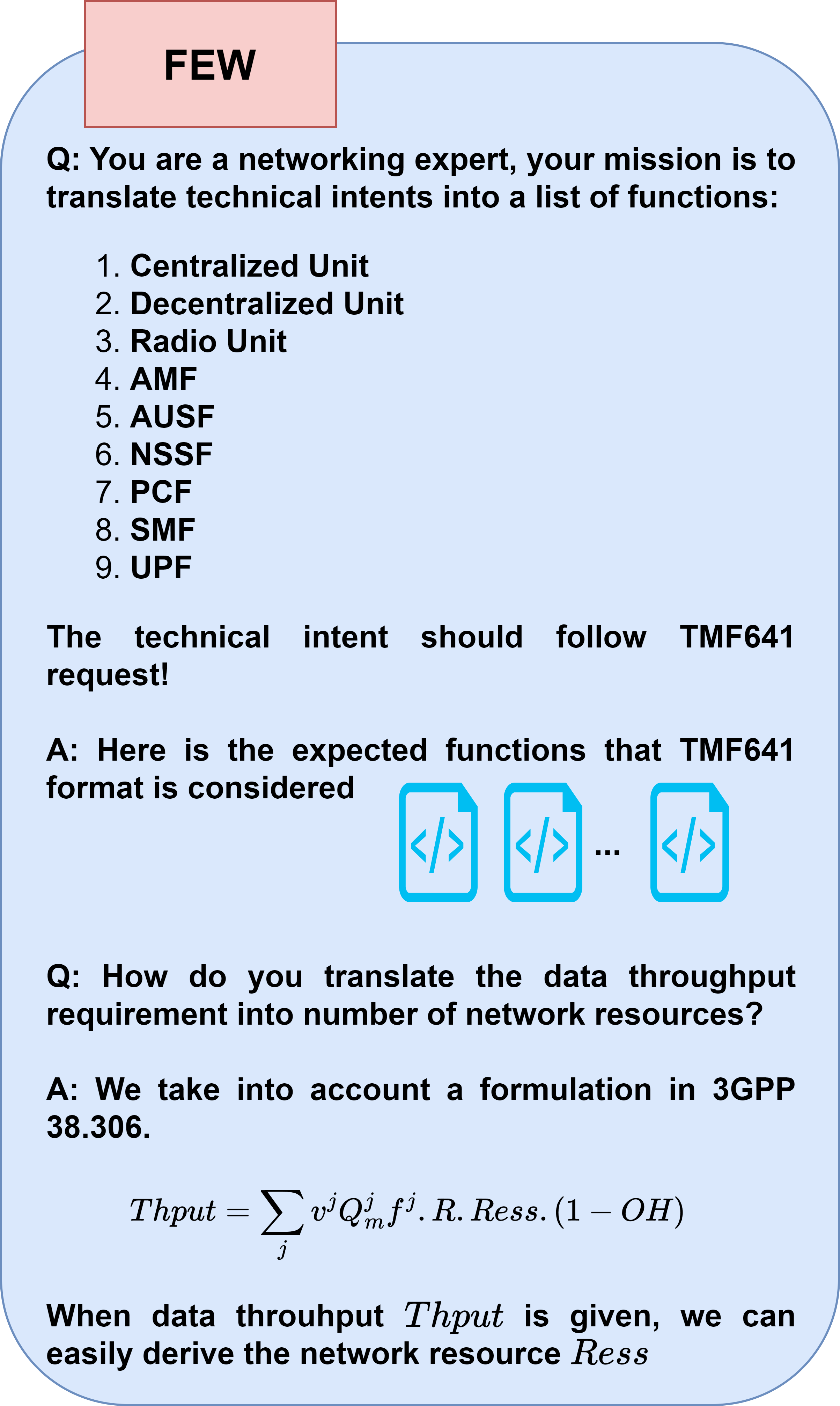}
    \caption{Few-shot prompting (FEW) }
    \label{fig:prompt2}
\end{figure}

In one-shot and few-shot prompting, we supply additional examples to the \acrshort{llm} models, specifying the service order given in JSON file and the expected results in YAML format. In particular, the one-shot prompt only contains the service order and the expected response, while few-shot CoT prompting additionally shows how to calculate certain fields in the expected response, given the technical intents in the service order as shown in Figure \ref{fig:prompt1} and Figure \ref{fig:prompt2}, respectively. Then, we pass the target service order to the \acrshort{llm} models and receive the generated results. Those obtained results are compared with a reference result, which is pre-designed by telecom expert, to evaluate how good/bad the translations are.

\subsection{LLM evaluation for intent translation/resolution - FEACI}

In order to evaluate how the answers generated from the \acrshort{llm} model,  general metrics such as BLEU \cite{papineniBLEUMethodAutomatic2002} and ROUGE \cite{linrouge2004} are commonly used in the literature. However, those metrics are mainly used to numerically examine the similarities between machine-translated text and references provided by telco experts. They have been shown to be inefficient to capture the semantic and context understanding of telecom-related responses where different words with analogous meaning may result in overall low scores. As a consequence, it is not appropriate to use them to evaluate the responses generated from \acrshort{llm} models in this work. To deal with this issue, we propose a novel evaluation metric, referred to as FEACI, to respectively evaluate the Format, Explanation, Accuracy, Cost, and Inference time of the generated responses as follows: 
\begin{itemize}
    \item Format: \textit{F-score}
    measures the correctness of the structure of the output proposed by the \acrshort{llm} compared to the reference structure of the RFS description (e.g., in JSON/YML, etc.). For each translation of the CFS-RFS pair, we consider a Boolean value: $F=1$ indicates that the generated network configuration is extractable by the resource orchestrator, otherwise $F=0$. Given a CFS catalog $\gls{D}$, the generated configuration of \acrshort{llm} $i$  is recorded in  the set $\mathcal{A}$. The format score of \acrshort{llm} $i$ (i.e., $0 \leq F_{\mathcal{A}}^{(i)} \leq 1$) is evaluated by the probability $\mathbb{P}$ as follows:
    \begin{equation}
        F_{\gls{D}}^{(i)} = \mathbb{P} \{F(o_a^i)=1 | o_a \in \mathcal{A} \}
    \end{equation}
  
    \item Explanation: \textit{E-score} indicates how good is an explanation that the \acrshort{llm} proposes for its output RFS. An explanation is not mandatory in our use case, and sometimes an explanation increases the number of tokens. However, for evaluation purposes, we consider the quality of the explanation to understand the result and improve the prompt, especially in cases where the \acrshort{llm} performs multiple calculations before providing the output (as described in the prompt of Figure \ref{fig:prompt2}). $E=1$ means \acrshort{llm} models can produce a satisfactory explanation for those responses generated, otherwise $E=0$. In this work, we applied human evaluation to assess the quality of the explanation. The explanation score of the generated set $\mathcal{A}$ of the \acrshort{llm} $i$ (i.e., $0 \leq E_{\mathcal{A}}^{(i)} \leq 1$), given CFS catalog $\gls{D}$ is as follows:
    \begin{equation}
        E_{\gls{D}}^{(i)} = \mathbb{P} \{E(o_a^i)=1 | o_a \in \mathcal{A} \}
    \end{equation}
    
    \item Accuracy: \textit{A-score} assesses the configuration values of \acrshort{llm} responses by comparing them to the respective reference values. Together with confirming the format of \acrshort{llm} responses, this metric aims to quantify the quality of \acrshort{llm} generation by statistically measuring the percentage of matching values between the response and the target reference.  
    In this sense, each output $o_a^i$ of \acrshort{llm} $i$ receives a score $A^{(i)} \in [0,1]$. The accuracy score of the generated set $\mathcal{A}$ of the \acrshort{llm} $i$ (i.e., $0 \leq A_{\mathcal{A}}^{(i)} \leq 1$), given CFS catalog $\gls{D}$ is measured by the following expectation $\mathbb{E}$:
    \begin{equation}
        A_{\gls{D}}^{(i)} = \mathbb{E}\{A^{(i)}(o_a^i)| o_a \in \mathcal{A}\} 
    \end{equation}
    \item Cost: \textit{C-score} measures how much we should pay (in terms of \textit{USD}) for the prompt tokens used. For the close-source \acrshort{llms} (i.e., Gemini 1.5 pro or GPT-4), the amount of input tokens (i.e., $N_{in}$) and output tokens  (i.e., $N_{out}$) are used to determined the cost (i.e., $c_i$ and $c_o$, respectively) 
  as 
    \begin{equation}
        C = c_i(N_{in}) + c_o(N_{out})
    \end{equation}
    With open-source \acrshort{llms} (i.e., LLama, Mistral, etc), there is no additional cost associated with the number of tokens generated (i.e., $C=0$). The normalized cost ($0\leq C^{(n)}\leq1$) is computed based on a threshold $C_0$ as follows:
    \begin{equation}
     C^{(n)}=\left\{\begin{matrix}
 \frac{C}{C_0} & \text{if $C \leq 10C_0$} \\
1 & \text{if $C > 10C_0$}
\end{matrix}\right.
\end{equation}
In this work, we consider the reference $C_o=0.1$ [USD]. Given CFS catalog \gls{D},  C-score is computed as follows:
\begin{equation}
    C_{\gls{D}}^{(i)} = \mathbb{E}\{C^{(n)}(o_a^i)| o_a \in \mathcal{A}\}
\end{equation}

    \item Inference time: \textit{I-score} measures the delay of generating the answer from the moment users query the \acrshort{llm} models until the responses are shown. Generally, this metric is heavily dependent on the processing time (i.e., $t_{proc}$) of input tokens (i.e.,$X_{inToken}$) in the transformer layer and time (i.e., $t_{gen}$) required for generating output tokens (i.e.,$Y_{outToken}$).
    \begin{equation}
        I = t_{proc}(X_{inToken}) + t_{gen}(Y_{outToken})
    \end{equation}
     The normalized version of this metric ($I^{(n)}$) is computed by defining a threshold $I_0$ ($I_0>0$) as follows:
     \begin{equation}
     I^{(n)}=\left\{\begin{matrix}
 \frac{I}{I_0} & \text{if $I \leq I_0$} \\
1 & \text{if $I > I_0$}
\end{matrix}\right.
\end{equation}
In this work, we consider the threshold $I_o=60$ [seconds]. Given CFS catalog \gls{D},  I-score is computed as follows:
\begin{equation}
    I_{\gls{D}}^{(i)} = \mathbb{E}\{I^{(n)}(o_a^i)| o_a \in \mathcal{A}\}
\end{equation}
\end{itemize}

The following evaluation score is proposed to examine the \acrshort{llm} models $i$ for the translation of service resolver from the data set \gls{D}:
\begin{equation}\label{eq:eval}
    \gls{Eserv}^{(i)} = \omega_1\times F_{\gls{D}}^{(i)} + \omega_2\times E_{\gls{D}}^{(i)} +\omega_3\times A_{\gls{D}}^{(i)} -\omega_4\times C_{\gls{D}}^{(i)} -\omega_5\times I_{\gls{D}}^{(i)}
\end{equation}
where $0\leq\omega_i\leq 1$, $i=1, \ldots, 5$, is the corresponding weight of each elementary score, representing the  weight of each individual contribution to the total score. 

\section{Results and discussion}\label{sec:results}

\subsection{Scenario description}
\begin{figure}[htp!]
    \centering
    \includegraphics[scale=0.5]{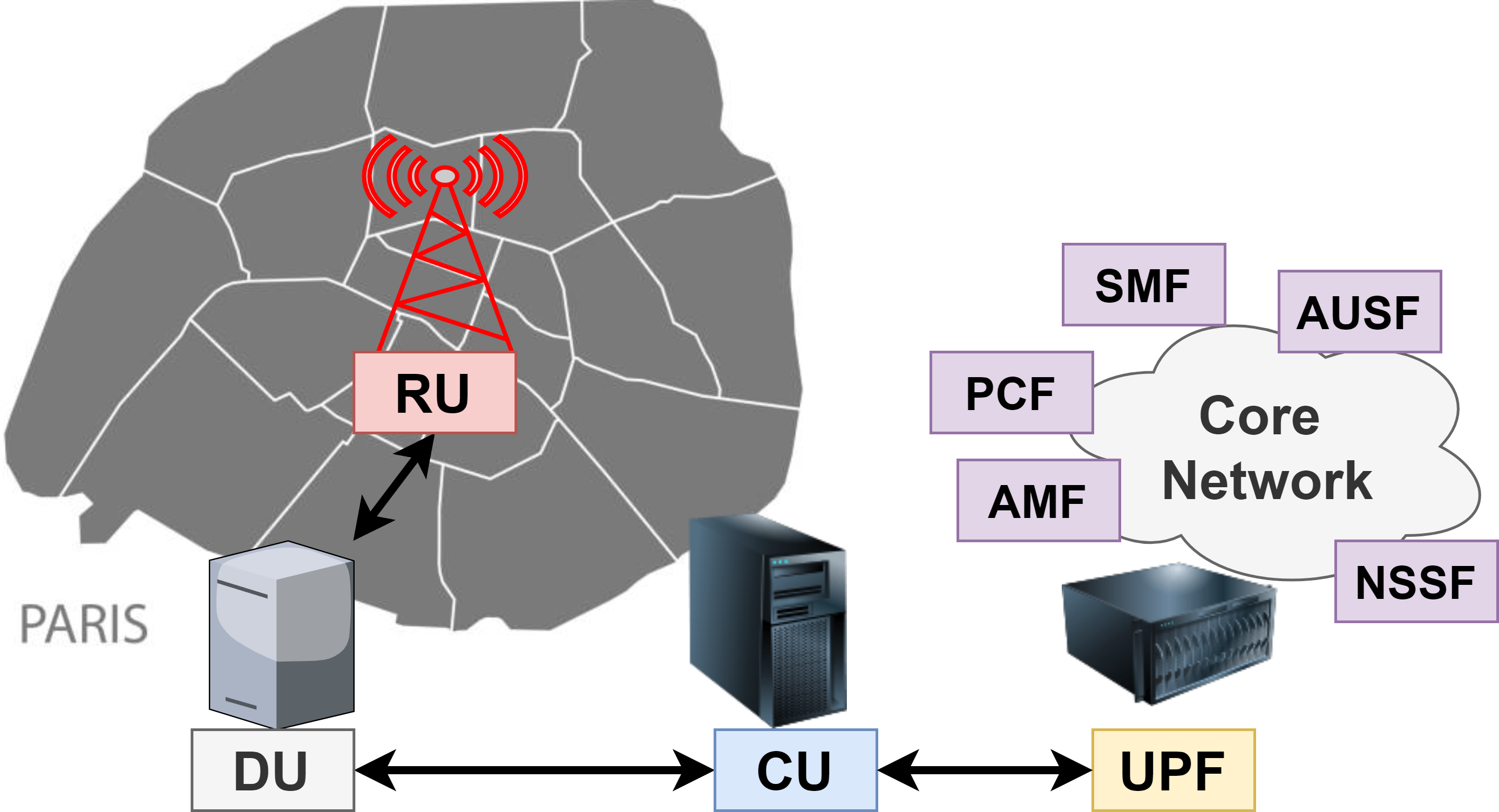}
    \caption{Network deployment for Parisien region}
    \label{fig:ran-core-paris}
\end{figure}

For the sake of simplicity, our \acrshort{e2e} network omits the transport network and comprises only two entities: \textit{(i)} a \acrshort{ran}
 and \textit{(ii)} a Core network. To establish an \acrshort{e2e} network service from user intents, the corresponding RAN and Core network functions must be appropriately configured, prior to the deployment. 
Figure \ref{fig:ran-core-paris} illustrates our target scenario in which a Radio Unit (RU), a Distributed Unit (DU) and a Control Unit (CU) are involved as the RAN functions, while User Plane Function (UPF), Access and Mobility Management Function (AMF), Policy Control Function (PCF), Session Manager Function (SMF), Authentication Server Function (AUSF) and Network Slice Selection Function (NSSF) are considered in the Core network. Customers express their demands through a User-friendly chatbot containing technical parameters (i.e., CFS) such as service slicing types, throughput requirements, maximum latency, numbers of supported users, etc. and wait for the deployment of their demands that are translated into a list of configurations for each network function (i.e., RFS). 

\subsection{Simulation setup}
Our experiment setup for testing open-source \acrshort{llm} models consists of a computing server, which is built based on 12xIntel(R) Xeon(R) E-2236 CPU of speed 3.4GHz and 2xNvidia H100 GPUs with 80GB of vRAM. We regard Gemini version 1.5 Pro (GEM 1.5) and ChatGPT version 4 (GPT-4) as closed-source models, since their underlying model specifications are inaccessible to the public. In this case, we use API to send the request to the closed-source model. On the other hand, the open models such as Llama-3 with 8B parameters (LLama I), 70B parameters (II), and Mistral models with 7B parameters (I), Mixtral 8x7B parameters (II) are also considered in our study. 

In general, closed-source models, which are only accessible via APIs, have more trainable parameters than open-source counterparts, which are locally stored in our computing server. We also demonstrate the inference cost of the \acrshort{llm} models by comparing the total cost associated with the number of input/output tokens required to tokenize the prompt and generate answers (i.e., in $USD$) per 1 million tokens used. In this regard, the costs associated with the open-source models are 0, in contrast to those found in the closed-source models.

Concerning the decoding strategy for the language model to generate the network configuration, temperature sampling  and nucleus sampling are used where temperature (i.e., $T$ parameter) and $top-p$ are  set at 0.2 and 0.9, respectively. Table \ref{tab:params} presents the parameters of open-source and closed-source \acrshort{llm} models employed in this work. 

\begin{table}[hbtp]
\centering
\caption{(Hyper)-Parameters of \acrshort{llm} models}
\label{tab:params}
\resizebox{\linewidth}{!}{%
\begin{tblr}{
  cells = {c},
  cell{1}{1} = {r=2}{},
  cell{1}{2} = {c=6}{},
  vlines,
  hline{1,4-10} = {-}{},
  hline{2-3} = {2-7}{},
}
\textbf{\textbf{Parameter(s)}} & \textbf{\textbf{\textbf{\textbf{Model(s)}}}} &           &              &              &               &                \\
                               & {GEM \\1.5}                                  & {GPT\\4}  & {Llama \\I}  & {LLama \\II} & {Mistral \\I} & {Mixtral \\II} \\
N-params                       & 1500B                                        & ~ 1760B~~ & ~ ~ ~8B~ ~ ~ & 70B          & 7B            & 45B            \\
Open                           & N                                            & N         & Y            & Y            & Y             & Y              \\
{Attention \\heads}            &       N/A                                       &     N/A      & 32           & 32           & 32            & 32             \\
{Token \\Size}                 & N/A                                         & N/A      & 128K         & 128K         & 32K           & 32K            \\
{Attention \\layer(s)}         &    N/A                                          & N/A       & 32           & 80           & 32            & 32             \\
{Activation \\function}        &     N/A                                         &     N/A      & SILU         & SILU         & SILU          & SILU           \\
{Cost In/Out\\(per 1M TK)}           &      1.25/5                                        &    10/30       & 0/0            & 0/0            & 0/0             & 0/0              
\end{tblr}
}
\end{table}





\subsection{Numerical results}

To obtain the statistical results of intent translation, we consider a catalog containing 10 network service orders that correspond to 10 \acrshort{cfs}s. Then, each translation from an \acrshort{cfs} into an \acrshort{rfs} is iterated 10 times for the average results. 

Table \ref{tab:bleu_tab} gives the BLEU/ROUGE performance of intent translation when zero-shot (ZERO), one-shot (ONE) and few-shot (FEW) prompting are applied on the closed-source/open-source \acrshort{llm} models. In particular, the generation of network configurations for each model and each prompting approach is compared to a reference configuration, which is pre-designed by our telco experts. As shown in this table, without prompts (i.e., ZERO), the scores of BLEU/ROUGE are 0 which signifies that the \acrshort{llm} models are not able to properly export the network configurations corresponding to the expert's expectation. When one-shot/few-shot prompts are applied to the models, they are capable of producing the correct format of network configuration. This explains why the scores improve when compared to those of  zero-shot prompting. However, those numbers do not have any significative meaning as we do not know how effective the translations are. 
As a consequence, we cannot use this general metric as a benchmark to compare the translation performance of the \acrshort{llm} models.

\begin{table}[hbtp]
\centering
\caption{Performance of intent translation with general metrics.}
\label{tab:bleu_tab}
\resizebox{\linewidth}{!}{%
\begin{tblr}{
  cells = {c},
  cell{3}{1} = {r=4}{},
  cell{7}{1} = {r=4}{},
  vlines,
  hline{1-3,7,11} = {-}{},
  hline{4-6,8-10} = {2-8}{},
}
     \textbf{Prompt}         & \textbf{Metrics} & \textbf{GEM} & \textbf{GPT~} & \textbf{LLAMA-I} & \textbf{LLAMA-II} & \textbf{Mistral-I} & \textbf{Mistral-II} \\
\textbf{ZERO} &       BLEU/ROUGE           &      0.0        &       0.0        &        0.0          &     0.0              &          0.0          &    0.0                 \\
\textbf{ONE}  & BLEU             &  0.415           &  0.382            &     0.236            &     0.331              &                    0.332&  0.361                   \\
              & ROUGE-1          &   0.164           &   0.177            &   0.124               &     0.173              &                   0.145 &       0.147             \\
              & ROUGE-2          &   0.130           &  0.145             &     0.097             &     0.115             & 0.111 &   0.111                  \\
              & ROUGE-L          &  0.164            &  0.177             &    0.124              &    0.173               &                   0.145 &      0.147               \\
\textbf{FEW}  & BLEU             &   0.455           &   0.291            &     0.376             &   0.429               &                    0.383&  0.394                  \\
              & ROUGE-1          &   0.243           &  0.117             &    0.599              &   0.158                &                   0.145 &   0.141                  \\
              & ROUGE-2          &  0.217            &  0.085            &   0.0               &    0.119               &                   0.111 &      0.106               \\
              & ROUGE-L          &  0.243            & 0.117              &   0.599               &    0.158               &                   0.145 &   0.141                  
\end{tblr}
}
\end{table}

Table \ref{tab:metric_tab} compares translation performance  when our metrics (i.e., FEACI) are used.  They examine the  Format (F), Explanation (E), Accuracy (A), normalized Cost (C), and normalized Inference time (I) of each language model to produce technical intents for RAN and Core networks. It shows that one-shot prompts is essential for the models to understand the output format of the network specifications, while few-shot prompting plays a significant role in reasoning why the certain results are generated (i.e., better scores obtained for the explanation (E) and Accuracy (A) metrics).

\begin{table}[hbtp]
\centering
\caption{Performance of intent translation with our metrics}
\label{tab:metric_tab}
\resizebox{\linewidth}{!}{%
\begin{tblr}{
  column{3} = {c},
  column{4} = {c},
  column{5} = {c},
  column{6} = {c},
  column{7} = {c},
  column{8} = {c},
  cell{1}{2} = {c},
  cell{2}{1} = {r=3}{},
  cell{5}{1} = {r=3}{},
  cell{8}{1} = {r=3}{},
  cell{11}{1} = {r=3}{},
  cell{14}{1} = {r=3}{},
  vlines,
  hline{1-2,5,8,11,14,17} = {-}{},
  hline{3-4,6-7,9-10,12-13,15-16} = {2-8}{},
}
\textbf{\textbf{Metrics}} & \textbf{Prompt} & \textbf{GEM} & \textbf{GPT~} & {\textbf{LLAMA}\\\textbf{-I}} & {\textbf{LLAMA}\\\textbf{-II}} & {\textbf{Mistral}\\\textbf{-I}} & {\textbf{Mistral}\\\textbf{-II}} \\
Format                    & ZERO            & 0.2          & 0.1           & 0.0                           & 0.0                            & 0.0                             & 0.0                              \\
                          & ONE             & 0.9          & 0.86          & 0.65                          & 0.78                           & 0.85                            & 0.87                             \\
                          & FEW             & 0.94         & 0.89          & 0.75                          & 0.82                           & 0.84                            & 0.88                             \\
Explain                   & ZERO            & 0.0          & 0.0           & 0.0                           & 0.0                            & 0.0                             & 0.0                              \\
                          & ONE             & 0.83         & 0.64          & 0.72                             & 0.75                              & 0.71                               & 0.86                                \\
                          & FEW             & 0.97         & 0.87          & 0.82                          & 0.86                           & 0.83                            & 0.89                             \\
Accuracy~ ~ ~ ~           & ZERO            & 0.0          & 0.0           & 0.0                           & 0.0                            & 0.0                             & 0.0                              \\
                          & ONE             & 0.84         & 0.54          & 0.72                          & 0.71                           & 0.78                           & 0.82                             \\
                          & FEW             & 0.93         & 0.62          & 0.82                          & 0.75                           & 0.84                            & 0.86                             \\
{Normalized\\Cost}        & ZERO            & 0.03         & 0.87           & 0.0                           & 0.0                            & 0.0                             & 0.0                              \\
                          & ONE             & 0.08         & 1.0           & 0.0                           & 0.0                            & 0.0                             & 0.0                              \\
                          & FEW             & 0.1         & 1.0           & 0.0                           & 0.0                            & 0.0                             & 0.0                              \\
{Normalized\\Inference}   & ZERO            & 0.24        & 0.36          & 0.62                          & 1.0                            & 0.65                            & 0.84                               \\
                          & ONE             & 0.27        & 0.41          & 0.69                          & 1.0                            & 0.67                            & 1.0                               \\
                          & FEW             & 0.31        & 0.45          & 0.71                          & 1.0                            & 0.73                            & 1.0                             
\end{tblr}
}
\end{table}

In terms of accuracy, zero-shot prompting (ZERO) is insufficient for the models to produce the expected results when there is no matching between the configuration file generated from the models and the references. One-shot prompting (ONE) considerably improves the accuracy to 54$\%$-84$\%$, when an example of format and expected results is given to the \acrshort{llm} models. Few-shot prompting (FEW) further enhances accuracy of the intent translation compared to one-shot prompting (ONE). Besides, the numerical values show that open-source \acrshort{llm} models allow us to achieve  accuracy performance similar to the one of  closed-source models, which are much larger and computationally expensive. 

In terms of  computational cost,  only closed-source models are taken into account as there is a price associated with the number of input/output tokens used to tokenize the user demands and generate the responses. Among them, the most expensive model is GPT-4. Furthermore, more examples we provide to the models, higher number of input tokens are used and turns out in higher cost. It explains why the cost related to the few-shot example is higher than the other prompting schemes. 

Finally, we compare how much (inference) time is needed for the model to generate the responses. Overall, this value depends on how many tokens that the models use to generate the results and is normalized by $I_o=60$ [seconds]. As we can see from the table, few-shot prompting requires longer inference time to generate the answers because it takes more time to process the additional tokens. In general, inference time of the Google Gemini 1.5 pro model is shortest when compared with the GPT v4 model and open-source model. Besides, inference time of the open-source models with higher number of parameters (i.e., Mistral-II and LLama-II) is longer  than with the lighter ones (i.e., Mistral-I and LLama-I).   

Table \ref{tab:tokens_tab} details the count of tokens that are consumed by  each language model to process the input prompt (i.e., zero-shot, one-shot, and few-shot) and to generate the response (output). In general, providing more examples through few-shot prompting enables the model to gain a deeper understanding of the context and produce more effective responses. As a result, a greater number of tokens are utilized during the inference process, thereby increasing the inference time overall. In contrast, zero-shot promptings do not require relevant examples to process user demands and results in lower number of tokens being used. In this scenario, while the inference time is significantly reduced when compared to few-shot prompting, the generated results are susceptible to \textit{hallucination} when the accuracy, the format, and explainability of the results do not meet user expectation.  

\begin{table}[hbtp]
\centering
\caption{Number of average input/output tokens used}
\label{tab:tokens_tab}
\resizebox{\linewidth}{!}{%
\begin{tabular}{|l|c|c|c|c|c|c|} 
\hline
              & \textbf{GEM} & \textbf{GPT~} & \textbf{LLAMA-I} & \textbf{LLAMA-II} & \textbf{Mistral-I} & \textbf{Mistral-II}  \\ 
\hline
\textbf{ZERO} & 2817/2614    & 2588/2036      & 2443/3401        & 2680/3164         & 3094/2646          & 3094/2234            \\ 
\hline
\textbf{ONE}  & 6511/3299    & 6858/2659      & 5697/4396        & 6723/3523         & 7328/1813          & 7328/2413            \\ 
\hline
\textbf{FEW}  & 8729/2683    & 8090/1865     & 7707/3386        & 8056/2520         & 9769/2486          & 9769/1944            \\
\hline
\end{tabular}
}
\end{table}

\hfill 

\begin{figure}[htp!]
    \centering
    \includegraphics[scale=0.45]{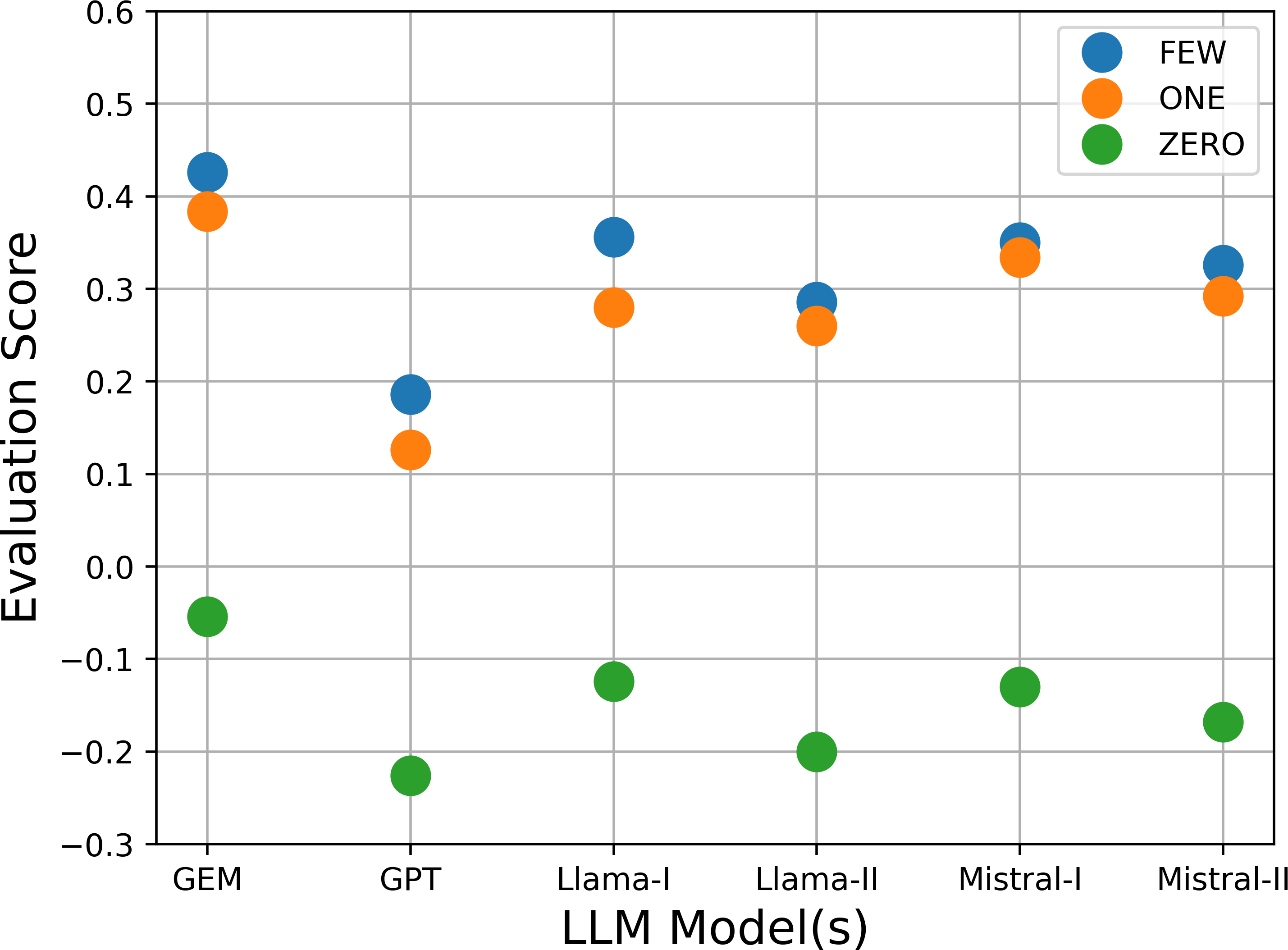}
    \caption{Evaluation scores of \acrshort{llm} models including the FEACI metrics}
    \label{fig:eval_score}
\end{figure}

Figure \ref{fig:eval_score} illustrates the total evaluation score, which is described in Equation \ref{eq:eval}, when the weights are evenly distributed (i.e., $\omega_i=0.2$ for $i =1,...,5$ ). It should be highlighted that the cost (C) and inference time (I) are normalized in this case. As shown in this figure, the evaluation score of the LLM translation when few-shot prompts are applied is higher than that of the one-shot and zero-shot prompts, when both accuracy and format scores are higher. GPT-4 model tends to perform the worst because the cost-related to the token consummation is the highest, whilst the accuracy, format, explanation scores are not better than the other models.


\section{Conclusion}\label{sec:conclusion}
In this study, we have presented an end-to-end network intent management framework that relies on \acrshort{llms}. Leveraging the capacity of \acrshort{llms} to comprehend complex tasks via prompt-based instruction, we have addressed the obstacles encountered in the translation and resolution phases of \acrshort{ibn}. Benchmarking the performance of both closed-source models (i.e., Google Gemini 1.5 pro, ChatGPT-4) and open-source models (i.e., LLama-8B(I), LLama-70B(II), Mistral-7B(I) and Mixtral-8x7B(II)) with few-shot (FEW), one-shot (ONE) and zero-shot (ZERO) prompts, we have demonstrated that prompting is an efficient method to substantially enhance the performance. Furthermore, we have proposed FEACI, a novel evaluation metric that assesses the performance of \acrshort{llms} in generating network configurations from user intents. The results reveal that open-source models can achieve similar or even better performance than closed-source models. 
Future studies should examine the techniques to further improve the reasoning capabilities and evaluation scores of the models.


%






\bibliographystyle{IEEEtran}

\IEEEtriggeratref{8}


%



\flushend

\end{document}

%% file: glossary.tex
\newacronym{cdf}{CDF}{Cumulative Distribution Function}
\newacronym{sla}{SLA}{Service Level Agreement}
\newacronym{api}{API}{Application Programming Interface}
\newacronym{ns3}{NS3}{Network Simulator 3}
\newacronym{sls}{SLS}{System Level Simulation}
\newacronym{e2ap}{E2AP}{E2 Application Protocol}
\newacronym{e2sm}{E2SM}{E2 Service Model}
\newacronym{sdn}{SDN}{Software-Defined-Networking}
\newacronym{ibn}{IBN}{Intent-Based Networking}
\newacronym{e2e}{E2E}{End-to-End}
\newacronym{qos}{QoS}{Quality-Of-Service}
\newacronym{embb}{eMBB}{enhanced-Mobible BroadBand}
\newacronym{urllc}{URLLC}{Ultra-Reliable and Low-Latency Communication}
\newacronym{ru}{RU}{Radio Unit}
\newacronym{du}{DU}{Distributed Unit}
\newacronym{cu}{CU}{Centralized Unit}
\newacronym{ran}{RAN}{Radio Access network}
\newacronym{nfvo}{NFVO}{Network Function Virtualization Orchestrator}
\newacronym{cfs}{CFS}{Custom-Facing Service}
\newacronym{rfs}{RFS}{Resource-Facing Service}
\newacronym{nro}{NRO}{Network Resource Orchestrator}
\newacronym{cpu}{CPU}{Central Processing Unit}
\newacronym{gpu}{GPU}{Graphics Processing Unit}
\newacronym{cuda}{CUDA}{Compute Unified Device Architecture}
\newacronym{ml}{ML}{Machine Learning}
\newacronym{drl}{DRL}{Deep Reinforcement Learning}
\newacronym{dl}{DL}{Deep Learning}
\newacronym{rl}{RL}{Reinforcement Learning}
\newacronym{llm}{LLM}{Large Language Model}
\newacronym{llms}{LLMs}{Large Language Models}
\newacronym{nlp}{NLP}{Natural Language Processing}
\newacronym{rnn}{RNN}{Recurrent Neural Networks}
\newglossaryentry{D}{%
name=\ensuremath{\mathcal{D}},
description={The product catalog  CFS }
}

\newglossaryentry{FD}{%
name=\ensuremath{F_\mathcal{D}},
description={Format score for the translation of \gls{D} }
}

\newglossaryentry{ED}{%
name=\ensuremath{E_\mathcal{D}},
description={Explanation score for the translation of \gls{D} }
}

\newglossaryentry{AD}{%
name=\ensuremath{A_\mathcal{D}},
description={Accuracy score for the translation of \gls{D} }
}

\newglossaryentry{CD}{%
name=\ensuremath{C_\mathcal{D}},
description={Cost score for the translation of \gls{D} }
}

\newglossaryentry{ID}{%
name=\ensuremath{I_\mathcal{D}},
description={Inference score for the translation of \gls{D} }
}

\newglossaryentry{A}{%
name=\ensuremath{\mathcal{A}},
description={The generated configuration of product catalog \gls{D} }
}

\newglossaryentry{omg}{%
name=\ensuremath{\omega_i},
description={Weight factor for FEACI metric}
}

\newglossaryentry{Eserv}{%
name=\ensuremath{\mathcal{E}_{serv}},
description={Evaluation score in service resolver}
}

\newglossaryentry{Ebuss}{%
name=\ensuremath{\mathcal{E}_{buss}},
description={Evaluation score in business resolver}
}

\newglossaryentry{alati}{%
name=\ensuremath{A_{lat_i}},
description={lower bound of confidence interval for latency }
}

\newglossaryentry{blati}{%
name=\ensuremath{B_{lat_i}},
description={upper bound of confidence interval for latency }
}

\newglossaryentry{llat}{%
name=\ensuremath{L_{lat}},
description={lower bound of confidence interval for latency }
}

\newglossaryentry{ulat}{%
name=\ensuremath{U_{lat}},
description={upper bound of confidence interval for latency }
}

\newglossaryentry{zlati}{%
name=\ensuremath{Z_{lat_i}},
description={latency value generated by the LLM model }
}